
\input harvmac
\input epsf

\def\nl{\hfil\break}
\def\tit{\bigskip\centerline{\it Department of Physics}
\centerline{\it Tokyo Institute of Technology}
\centerline{\it Oh-okayama, Meguro-ku}
\centerline{\it Tokyo, JAPAN  152}
\vskip .3in}

\def\nvc{{ \sl Nuovo Cim. }}
\def\npb{{ \sl Nucl. Phys. }}
\def\prc{{ \sl Phys. Rep. }}
\def\prd{{ \sl Phys. Rev. }}
\def\prl{{ \sl Phys. Rev. Lett. }}
\def\plb{{ \sl Phys. Lett. }}
\def\rmp{{ \sl Rev. Mod. Phys. }}

\font\cmss=cmss10 \font\cmsss=cmss10 at 7pt
\def\IZ{\relax\ifmmode\mathchoice
{\hbox{\cmss Z\kern-.4em Z}}{\hbox{\cmss Z\kern-.4em Z}}
{\lower.9pt\hbox{\cmsss Z\kern-.4em Z}}
{\lower1.2pt\hbox{\cmsss Z\kern-.4em Z}}\else{\cmss Z\kern-.4em Z}\fi}
\def\undertext#1{\vtop{\hbox{#1}\kern 1pt \hrule}}
\def\half{{1\over2}}
\def\c#1{{\cal{#1}}}
\def\slash#1{\hbox{{$#1$}\kern-0.5em\raise-0.1ex\hbox{/}}}
\def\dirac{\hbox{$\partial$\kern-0.5em\raise0.3ex\hbox{/}}}
\def\dslash{\hbox{$\partial$\kern-0.5em\raise0.3ex\hbox{/}}}
\def\Dirac{\hbox{{\it D}\kern-0.52em\raise0.3ex\hbox{/}}}

\def\kslash{\hbox{{\mit k}\kern-0.4em\raise0.3ex\hbox{/}}}
\def\pslash{\hbox{{\it p}\kern-0.48em\raise-0.3ex\hbox{/}}}
\def\qslash{\hbox{{\mit q}\kern-0.5em\raise-0.3ex\hbox{/}}}
\def\gsim{\mathrel{\raise.3ex\hbox{$>$\kern-.75em\lower1ex\hbox{$\sim$}}}}
\def\lesssim{\mathrel{\raise.3ex\hbox{$<$\kern-.75em\lower1ex\hbox{$\sim$}}}}

\overfullrule=0pt
\def\nf{N_F}
\def\drho{\delta\rho}

\def\hm{\hat m}
\def\mtp{m_{t,pert}^2}
\def\gf{G_F}
\def\zzero{Z_{\chi^0}}
\def\zplus{Z_{\chi^+}}

\def\thw{\theta_W}

\def\QL{Q_L}
\def\TR{T_R}
\def\TL{T_L}
\def\TTL{U_L}
\def\al#1{\alpha_{#1}}
\def\ss#1{{\scriptscriptstyle#1}}

\def\tql{\tilde q_{\ss L}}
\def\ql{q_{\ss L}}
\def\tbl{\tilde b_{\ss L}}
\def\bl{b_{\ss L}}
\def\Fl{F_{q_L}}
\def\striv{s_{triv}}
\def\tr{t_{\ss R}}
\def\ttr{\tilde\tr}
\def\Fr{F_{t_R}}
\def\tp{\tilde \phi}
\def\Fp{F_\phi}
\def\lsb{\c L_{sb}}
\def\mq{m_{\tilde q}}
\def\mt{m_{\tilde t}}
\def\tl{t_{\ss L}}
\def\ttl{\tilde\tl}
\def\Ftl{F_{\tl}}
\def\d{\partial}
\def\sumo{\sum_{i=1}^{\nf}}
\def\lkin{\c L_{kin}}
\def\lpot{\c L_{pot}}
\def\Fpi{F_{\phi_i}}
\def\aoneb{A_{1,bare}}
\def\atwob{A_{2,bare}}
\def\athreeb{A_{3,bare}}
\def\aone{A_1}
\def\atwo{A_2}
\def\athree{A_3}
\def\szero{s_0}
\def\sb{s_{bare}}

\def\zz{Z_{\chi^0}}
\def\zp{Z_{\chi^+}}
\def\chip{\chi^+}
\def\chiz{\chi^0}

\def\intk{\int_{k^2<\Lambda^2}\! {d^4k\over(2\pi)^4}}

\def\yren{y^2(\szero)}
\nopagenumbers\abstractfont\hsize=\hstitle\rightline{TIT/HEP--233}
\vskip 1in\centerline{\titlefont
Effects of new physics to the the $\rho$--parameter}
\centerline{\titlefont
in the supersymmetric standard model}
\abstractfont\vskip .5in\pageno=0
\centerline{Kenichiro Aoki\footnote{$^\star$}{\hskip-1mm%
email:{\tt~ken@phys.titech.ac.jp}}}
\baselineskip=11.5pt plus 2pt minus 1pt
\tit
\medskip
\centerline{\bf Abstract}
The contribution to the $\rho$--parameter
of the quark--Higgs sector of the
supersymmetric standard model is computed
non--perturbatively using the large--$\nf$ expansion
for the case tan$\beta$=0.
An explicit formula is found for the $\rho$--parameter
which is ill-defined unless the triviality cutoff
is taken into account.
The cutoff dependence of the $\rho$--parameter
is found to be large if and only if the top mass is larger
than $2v$ so that the cutoff scale
is of the order of the electroweak scale.
These non--universal effects are the reflections  at low energies
of the physics beyond the supersymmetric standard model.
By also considering the effects of the
soft--breaking terms, cutoff effects in both decoupling
and non--decoupling effects are analyzed.

\Date{8/93}
\newsec{Introduction}
The supersymmetric standard model is a natural generalization
of the standard electroweak model that is perhaps realized
in nature\ref\SSM{For some of the reviews on the subject, see\nl
P. Fayet, S. Ferrara, \prc{\bf 32} (1977) 334\nl
J. Wess, J. Bagger, {\sl ``Supersymmetry and supergravity",}
Princeton University Press (1983)\nl
S.J. Gates, M.T. Grisaru, M. Ro\v cek, W. Siegel, {\sl ``Superspace or
one thousand and one lessons in supersymmetry"}, Benjamin/Cummings, (1983)\nl
H.P. Nilles, \prc{\bf 110} (1984) 1\nl
H.E. Haber, G.L. Kane, \prc{\bf117} (1985) 75}.
However, to this date, no supersymmetric partners of
ordinary particles have been observed
and they are assumed to be heavy enough to have avoided detection so far.
When the mass of a particle is generated by a
coupling constant,
as is the case in the (supersymmetric) standard model,
the decoupling theorem of \ref\AC{T. Appelquist, J. Carazzone,
\prd{\bf D11} (1975) 2856} does not apply and
the physical effects of this particle need not decrease
as the particle becomes heavier.
Using this fact, some important
perturbative restrictions have been placed
on yet unseen particles in the (supersymmetric) standard model,
such as the top
\ref\RADIATIVE{For
reviews on
radiative corrections in the standard model, see for instance,\nl
K.~Aoki, Z.~Hioki, R.~Kawabe, M.~Konuma, T.~Muta, {\sl Suppl.
Progr. Theor. Phys.,} {\bf 73} (1982) 1\nl
M.~B\"ohm, H.~Spiesberger, W.~Hollik, {\sl Fortshr. Phys. }{\bf 34}
(1986) 687\nl
M. Consoli, W. Hollik, F. Jegerlehner, in {\sl ``LEP Physics Workshop''}
(1989)\nl
M. Peskin, Lectures given at the 17th SLAC summer institute (1989)\nl
R.D. Peccei, Lectures given at the Beyond the standard model I$\!$I
symposium (1989)\nl
P.~Langacker, M.~Luo, A.~Mann, \rmp{\bf64} (1992) 87}%
\ref\SRHO{L. Alvarez--Gaum\'e, J. Polchinski,
M. Wise, \npb{\bf 221B} (1983) 435\nl
R. Barbieri, L. Maiani, \npb{\bf B224} (1983) 32\nl
C.S. Lim, T. Inami, N. Sakai, \prd{ \bf 29D} (1984) 1488}%
\ref\SRHOR{A. Bilal, J. Ellis, G.L. Fogli, \plb{\bf 246B} (1990) 459\nl
P. Gosdzinsky, J. Sol\`a, \plb{\bf 254B} (1991) 139\nl
and references therein.}.
As the coupling constant becomes stronger, the perturbation theory
becomes less reliable so that one also needs to analyze the
behavior of non--decoupling effects outside the perturbative regime.

This problem raises a number of conceptually interesting questions:
The $\rho$ parameter increases with the fermion mass; what
happens when we keep on increasing the fermion mass?
Does it keep on increasing or somehow saturate?
Is the perturbative bound on the top mass legitimate
when we also consider the non--perturbative regime?
In which region is the perturbation theory valid?
What are the non--perturbative effects to the
$\rho$ parameter, or aren't there any?
In the standard model, these questions were answered
using the $1/\nf$ expansion in \ref\AP{K. Aoki, S. Peris,
UCLA preprint, /UCLA/92/TEP/23 (1992)}.
(For the Higgs case, also see \ref\CPP{S. Cortese,
E. Pallante, R. Petronzio, \plb{\bf 301B} (1993) 203},
in regards to the $S$ parameter, see \ref\APS{
K. Aoki, S. Peris, \prl{\bf 70} (1993) 1743}.)
First, we need to remember that the theory has a
physical cutoff, namely the triviality scale, and that
the fermion mass
can only be at most a few times the symmetry breaking
scale \ref\LATTICE{J.~Shigemitsu, \plb{\bf226B} (1989) 364\nl
I-H.~Lee, J.~Shigemitsu, R.~Shrock, \npb{\bf B330} (1990) 225,
\npb{\bf B335} (1990) 265\nl
W.~Bock, A.K.~De, C.~Frick, K.~Jansen, T.~Trappenburg,
\npb{\bf B371}~(1992)~683\nl
S.~Aoki, J.~Shigemitsu, J.~Sloan, \npb{\bf B372} (1992) 361\nl
Proceedings of the International Symposium on Lattice Field Theory,
KEK (1991)\nl and references therein.}%
\ref\EG{M.B. Einhorn, G. Goldberg, \prl{\bf 57} (1986) 2115}%
\ref\ON{K. Aoki,  \prd{\bf D44} (1991) 1547}.
It was found that when  the fermion mass increases
to few times the vacuum expectation value ($\gsim2.5v$,
$v=246$ GeV), the
$\rho$--parameter depends substantially on how the
cutoff is applied in the theory.
This dependence is the sensitivity of the non--decoupling
effects, such as the $\rho$ parameter, to the physics
beyond the standard model.
In a renormalizable model, such as the standard model,
it is at first sight surprising that a low energy parameter
can depend on the details at high
energies and this fact is special to non--decoupling effects.

In this work, we compute the contribution to the $\rho$--parameter
from the supersymmetric Yukawa coupling non--perturbatively
using the $1/\nf$ expansion.
The effects of soft supersymmetric breaking terms will
be included.
In four dimensions, the only renormalizable
supersymmetric interactions that may give rise to non--decoupling
effects are the gauge and the Yukawa interactions.
The gauge couplings in the case of the (supersymmetric)
standard model have been measured to be small
so that the Yukawa--Higgs coupling is perhaps
of more interest.
We will use the supersymmetric Yukawa--Higgs sector
of the standard model solved
in the large--$\nf$ limit \ref\KASUSY{K. Aoki,
\prd{\bf D46} (1992) 1123}
to compute the $\rho$--parameter.
We believe that our explicit results using
the $1/N$ expansion  clearly delineates the
physics underlying the results.
In the supersymmetric case,
the formulation of supersymmetric  theories on the lattice
is still an open problem \ref\SUSYLAT{
H. Nicolai, P. Dondi, \nvc{\bf41A} (1977) 1\nl
T. Banks, P. Windey, \npb{\bf B198} (1982) 226\nl
J. Bartels, J.B. Bronzan, \prd{\bf D28} (1983) 818\nl
J. Bartels, G. Kramer,  {\sl Z. Phys.} {\bf C20} (1983) 159\nl
M.F.L. Golterman, D.N. Petcher, \npb {\bf 319} (1989) 307}
so that it might be difficult to obtain non--perturbative
information through other approaches.

In section 2, we explain precisely
how our calculation applies to
the supersymmetric
standard electroweak model.
We also explain how the large--$\nf$ limit is taken and how
the model is solved in this limit.
In section 4, we compute the $\rho$--parameter in the
case with no soft supersymmetry breaking terms and when the
model is supersymmetric.
This case is much simpler than the case including
the effects of the
soft breaking terms which is dealt with in section 5.
We conclude with a discussion of our results in section 6.
\newsec{The supersymmetric Yukawa--Higgs sector of the supersymmetric
standard model in the large--$\nf$ limit}
In this section, we briefly explain how the quark--Higgs (or the lepton--Higgs)
sector of the supersymmetric standard model is solved in the large--$\nf$
limit. We refer to \KASUSY\ for details.

Consider three scalar supermultiplets, $\QL, \TR$ and $\Phi$
which transform under \nl
SU$(\nf)_{\ss L}\times$U$(1)_{\ss L}\times$U$(1)_{\ss R}$ as
\eqn\transf{
\QL\mapsto^{i\al{\ss L}}\TTL\QL\qquad
  \Phi\mapsto e^{i(\al{\ss L}-\al{\ss R})}\TTL\Phi\qquad
\TR\mapsto e^{i\al{\ss R}}\TR
}
where $\TTL\in$SU$(\nf)$ and $\al{\ss L},\al{\ss R}$ are real.
The boson, fermion and the auxiliary field components of the
superfields $\QL,\TR$ and $\Phi$ are denoted
$(\tql,\ql,\Fl),(\ttr,\tr,\Fr)$ and $(\phi,\tp,\Fp)$
respectively.

The action of the model is derived from the standard kinetic
term, the superpotential
\eqn\spot{W=y\QL^\dagger\Phi\TR}
and the soft supersymmetry breaking terms in the Lagrangian
(in components)
\eqn\deflsb{
-\lsb=\mq^2\left|\tql\right|^2+\mt^2\left|\ttr\right|^2
}
The  model has global symmetry SU$(\nf)_{\ss L}\times$U$(1){\ss R}\times
$U$(1)_{\ss L}\times$U$(1)_{\c R}$
where the U$(1)_{\c R}$ \c R--symmetry acts on the fields as
\eqn\raction{
\tql\mapsto e^{i\varphi_{\c R}}\tql\qquad
\ttr\mapsto e^{i\varphi_{\c R}}\ttr\qquad
\tp\mapsto e^{-i\varphi_{\c R}}\tp\qquad
\Fl\mapsto e^{i\varphi_{\c R}}\Fl\qquad
\Fr\mapsto e^{i\varphi_{\c R}}\Fr
}
while keeping the other fields
fixed. The action of the rest of the
transformations was defined in \transf.
The soft supersymmetry breaking terms in \deflsb\
are the most general ones
while retaining this symmetry.
Another soft breaking term, $\tql\phi\ttr$ may be added
which would break the U$(1)_{\c R}$
${\c R}$--symmetry to $\IZ_2$ symmetry.
We shall not consider this term below.

In terms of components, the Lagrangian including
the auxiliary fields is
\eqn\totlagrangian{\eqalign{\c L&=\lkin+\lpot+\lsb\cr
-\lkin &= \left|\d\phi\right|^2+
\sumo\left|\d\tql^i\right|^2+\left|\d\ttr\right|^2
+\half\overline{\tp}\dirac\tp
+\overline\ql\dirac\ql+\overline\tr\dirac\tr\cr
&\qquad-\sumo\left|\Fpi\right|^2
-\sumo\left|\Fl^i\right|^2-\left|\Fr\right|^2\cr
-\lpot & = y\left[\overline\ql\phi\tr+
\tql^{\dagger}\overline{\tp}\tr+\overline\ql\tp\ttr
\right]
+y\biggl[\tql^{\dagger}\phi\Fr
+\tql^{\dagger}\Fp\ttr
+\Fl^{\dagger}\phi\ttr\biggr]+{\rm h.c.}\cr}}
where h.c. denotes the Hermitean conjugate terms.

The auxiliary fields may be integrated out to obtain  a more familiar
form of the Lagrangian
\eqn\fulllagrangian{\eqalign{
-\c L &=-\c L_{sb}+
 \left|\d\phi\right|^2+\sumo\left|\d\tql^i\right|^2
  +\left|\dslash\ttr\right|^2
  +\half\overline{\tp}\dslash\tp
   +\overline\ql\dslash\ql+\overline\tr\dslash\tr\cr
&\qquad  +y\left[\overline\ql\phi\tr+\tql^\dagger\overline{\tp}\tr
  +\overline\ql\tp\ttr+h.c.\right]
  +y^2\biggl[|\phi|^2|\ttr|^2
   +\sumo|\tql^i|^2|\ttr|^2+|\tql^\dagger\phi|^2\biggr]\cr
  }}

This model is a part of any supersymmetric generalization of the
standard electroweak model in the quark--Higgs or the lepton--Higgs
sector (for $\nf=2$).
Only one Yukawa coupling has been retained and the electroweak gauge
couplings have been set to zero.
To obtain the $\rho$--parameter to leading order in
the gauge coupling constants, $g,g'$, it is only necessary to
use external gauge fields.
In this model, only one Higgs supermultiplet
exists in the model so that there is no mixing of the Higgs
multiplets.
This case is often referred to as the $\tan\beta=0$ case.

The vacuum expectation values of the superfields are
\eqn\vevs{
\left\langle\Phi_i\right\rangle=v/\sqrt2\,\delta_{i1}\qquad
\left\langle\QL\right\rangle=\left\langle\TR\right\rangle=0
}
The vacuum expectation value spontaneously breaks the symmetry
of the model to SU$(\nf-1)\times$U$(1)_{L+R}\times$U$(1)_{\c R}$.

The large--$\nf$ limit is taken in this model by
fixing $v^2/\nf$, $y^2\nf$ while taking $\nf$
to infinity.
When no soft breaking terms are present, the leading
order quantum effects that are of the same order
as the classical terms may be
summarized in the supergraphs in \fig\figloop{}\nl
\centerline{\epsfysize=3.0cm\epsfbox{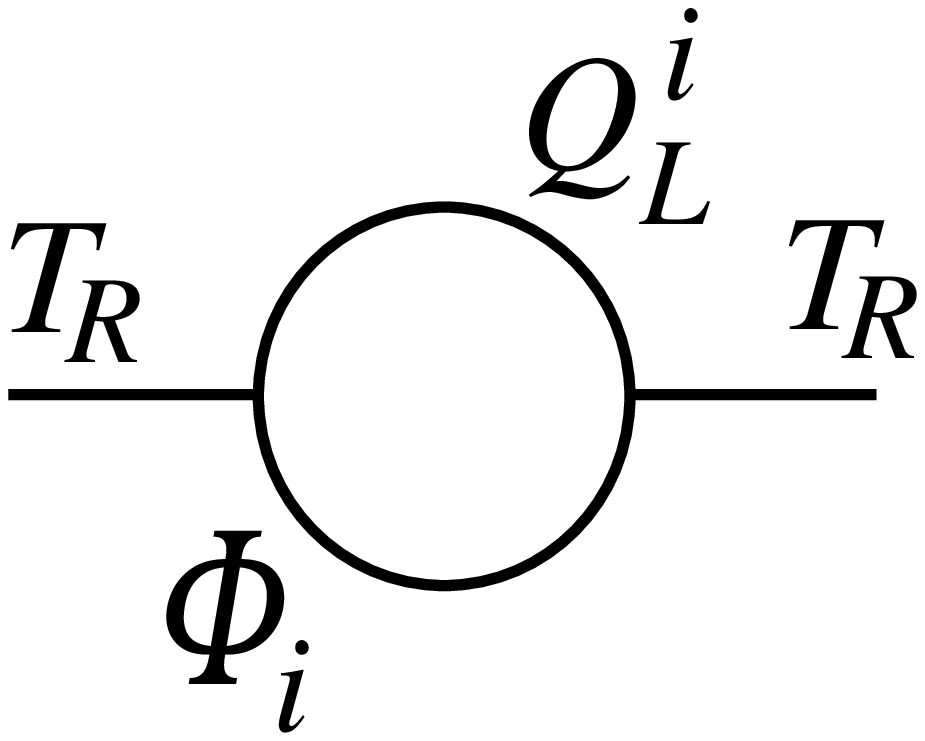}}
\smallskip\noindent
\figloop\ {\it Leading order contributions to the model when
no soft supersymmetry breaking terms are present.}
\medskip\noindent
The model is the natural supersymmetric generalization
of the O$(N)$ model \ref\ON{K.G.~Wilson, \prd{\bf D7} 2911\nl
L.~Dolan, R.~Jackiw, \prd{\bf D9} (1974) 3320\nl
H.J.~Schnitzer, \prd{\bf D10} (1974) 1800,2042\nl
S.~Coleman, R.~Jackiw, H.D.~Politzer, \prd{\bf D10} (1974) 2491\nl
  L.F.~Abbott, J.S.~Kang, H.J.~Schnitzer, \prd{\bf D13} (1976) 2212\nl
  W.A.~Bardeen, M.~Moshe, \prd{\bf D28} (1983) 1372\nl
  M.B.~Einhorn, \npb{\bf B246} (1984) 75, and references therein}.
Including the effects of the soft breaking terms, the leading order
effects may be neatly summarized in the following effective
Lagrangian in which only the kinetic terms
of the component fields in the $\TR$  multiplet receive
radiative corrections:
\eqn\eqleff{\eqalign{\c L^{eff} &= \c L^{eff}_{kin}+\lpot+\lsb\cr
  -\lkin^{eff} & = \aoneb\left|\d\ttr\right|^2+\atwob\overline\tr
  \dirac\tr-\athreeb\left|\Fr\right|^2+\hbox{\rm other terms.}\cr}
}
where
\eqn\eqas{\eqalign{
    \aoneb(p^2)  &
     =1-{y^2\nf\over(4\pi)^2}\ln{p^2\over\sb}
       \cr
    \atwob(p^2)
    &=1-{y^2\nf\over2(4\pi)^2}\left[\ln{p^2\over\sb}
      +\ln{\mq^2\over\sb}+\left(1+{\mq^2\over p^2}\right)^2\ln
          \left(1+{p^2\over\mq^2}\right)-{\mq^2\over p^2}\right]\cr
  \athreeb(p^2)
   &=1-{y^2\nf\over(4\pi)^2}\left[\ln{\mq^2\over\sb}
         +\left(1+{\mq^2\over p^2}\right)\ln
          \left(1+{p^2\over\mq^2}\right)\right]
         \cr}
}
in momentum space.
$\sb$ denotes a regulator dependent number of dimension
of mass squared whose explicit expression will not be necessary.
(In dimensional regularization, $\ln\sb=1/\epsilon-\gamma_{\ss E}
+2+\ln(4\pi\mu^2)$, where $\gamma_{\ss E}$ is the Euler--Mascheroni
constant and $\mu$ is the scale parameter introduced by dimensional
regularization.)
This characterization of the leading large--$\nf$ results
will prove to be useful in the subsequent computations.
When there are no soft breaking terms
$A_{i,bare}$'s are all equal since the fields $\ttr,\tr,\Fr$ all
belong to the same supermultiplet.

We renormalize the coupling constant $y$ at an arbitrary
momentum squared scale $\szero$ as
\eqn\eqren{
  \yren\nf = {y^2\nf\over1-y^2\nf/(4\pi)^2\ln\szero/\sb}
}
$\mq^2$ and $v^2$ need no renormalization to this order.
Strictly speaking, $\mt^2$ is also renormalized, but this
will not play a role in the following.

The full propagators that enter the  calculations
of physical quantities cease to make sense at some
energy scales, the smallest of which we
call the ``triviality scale''. The triviality
scale $\striv$ is determined through the equation
$\athreeb(\striv)=0$ and all particle masses need to
be below this scale for consistency.
$\striv$ is a physical scale independent of the
renormalization scheme.
Also, $\striv$ has the typical non--perturbative
dependence on the coupling; for instance, when
$\mq^2=\mt^2=0$, $\striv=\szero\exp((4\pi)^2/(\yren\nf))$.
The spectrum, which was arbitrary within perturbation
theory, is restricted by the consistency of the
theory.
In particular, the mass of the fermion $t$
can be at most a few times the scale $v$.
For the possible spectra depending on the soft--breaking parameters,
see \KASUSY.
\newsec{The $\rho$--parameter }
At energies well below the $W,Z$ gauge boson masses,
the interactions in the electroweak model
may be effectively described by the current current interaction
of the following form
\def\jcp{J_\mu^{+}}
\def\jcm{J_\mu^{-}}
\def\jnc{J_\mu^{0}}
\eqn\eqrhodef{-\c L_{JJ}
  ={\gf\over\sqrt2}\left(\jcp{\jcm}+\half\rho\,\jnc\jnc\right)}
Here, $J_\mu^\pm,\jnc$ denote the charged and the neutral weak
currents. $\rho$ represents the relative strength of
the neutral current interaction to the charged one.
The weak interactions are mediated by $W,Z$ gauge bosons whose
massive modes are the Nambu--Goldstone
bosons of the ungauged theory. Consequently,
the $\rho$--parameter may be
characterized using the effective Lagrangian for the
Nambu--Goldstone bosons $\chip,\chiz$ as \ref\LYTEL{
R.S.~Lytel, \prd{\bf 22D} (1980) 505
}
\eqn\rhong{\rho=\left.{\zp(p^2)\over\zz(p^2)}\right|_{p^2=0}}
using the effective Lagrangian
\eqn\nglag{-\c L_{eff}^{NG} =
\zplus\left|\d_\mu\chi^+-{gv\over2}W_\mu^+\right|^2
+\half\zzero\left(\d_\mu\chi^0-{gv\over2\cos\thw}Z_\mu\right)^2
+\hbox{ other terms}}
This formula characterizes the deviation of the $\rho$--parameter
from unity as the lack of custodial symmetry amongst the
Nambu--Goldstone bosons.
In the $1/\nf$ expansion, $\chip$ may be taken to be
any Nambu--Goldstone boson in the $\nf-1$ multiplet of
SU$(\nf-1)$ due to the residual symmetry.
These expressions \eqrhodef,\rhong\ are exact up to higher
order terms in the electroweak gauge coupling
constants and in the $1/\nf$ expansion.
\newsec{$\rho$--parameter in the case with no
soft breaking terms}
In this section, we compute the $\rho$--parameter
non--perturbatively in the case when $\mq^2=\mt^2=0$.
This case is simpler than the more general case
dealt with in the next section yet illustrates
some of the main concepts.
The leading order corrections to the gauge boson propagators
come from the graphs
of $\c O(1/\nf)$ in \fig\figtwo{}
in the component formulation.
 \centerline{\epsfxsize=\hsize\epsfbox{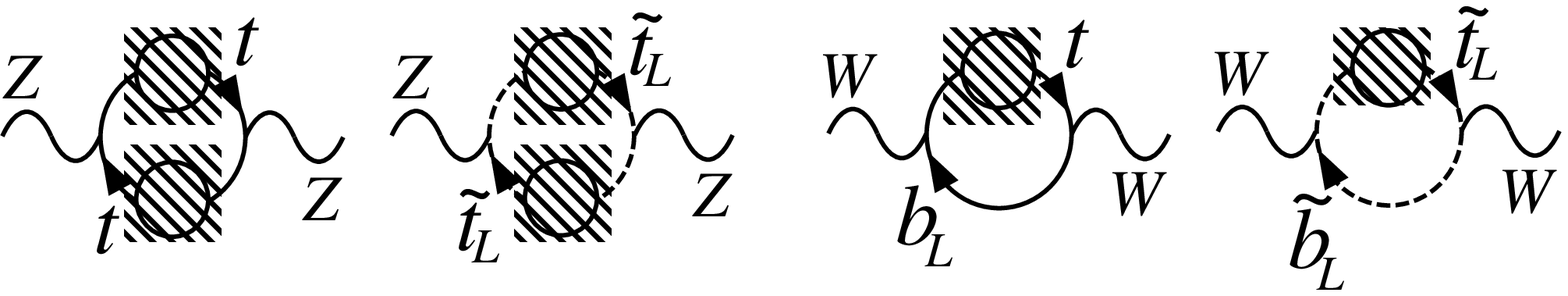}}
\smallskip\noindent
\figtwo{\it \ Leading order quantum corrections to the
$Z$, $W$ propagators.}\medskip\noindent
In the figure, we collectively denoted
$\ql^i\equiv\bl,\ \tql^i\equiv\tbl\ (i\neq1)\ $.
There are also seagull contributions of the same order,
however these contributions are the same for
both $W$ and $Z$ so that they
cancel in their contribution to the $\rho$ parameter and
we shall not discuss them here.

Using the full propagators in the large $\nf$ limit derived
from \eqleff\ and using the definition of the $\rho$
parameter in \eqrhodef, we arrive at the expression
for the $\rho$ parameter to leading order in the
$1/\nf$ expansion:
\eqn\rhosusy{
\drho
    ={2\hat m^4\over v^2}
  \intk{1\over k^2\left(\aone(k^2)k^2+\hat m^2\right)^2}}
We have used the renormalized parameters in this expression;
here, $\hat m\equiv y(\szero)v/\sqrt2$ and
$\aone$ is the same formula as $\aoneb$
with the replacement, $y^2\mapsto\yren,\ \sb\mapsto\szero$;
{\it i.e.}, $\aone(p^2)\equiv1-\yren\nf/(4\pi)^2\ln p^2/\szero$.
Since this is the contribution from one multiplet, we
need to multiply by the number of colors.
The renormalization scale $\szero$ is arbitrary.
Since we will only be interested in the behavior
of the $\rho$ parameter with respect to the
particle masses which are both physical quantities,
a change in the renormalization
scale does not affect the results at all.

The integrand in this expression for the $\rho$--parameter is
integrable both at $k^2=0$ and at $k^2=\infty$.
However, we cannot integrate over $k^2$ from zero to infinity due
to the existence of a pole in the integrand.
The location of this pole is always above the triviality scale
$\striv$ so that we cutoff this integral at the scale $\Lambda^2<\striv$.
This is also natural since the triviality scale is the
intrinsic cutoff of the theory,
above which scale the theory breaks down.
The $\rho$--parameter now depends on how the cutoff is implemented
and to understand the dependence of this parameter on the cutoff
procedure, we varied the cutoff scale as $\Lambda/\sqrt{\striv}=0.1,
0.5,0.9$.
The result is shown in \fig\figsusy{}\ as the
variation of the $\rho$--parameter against the
physical mass of the particles in the $\TL,\TR$ supermultiplets
which are degenerate in mass.
The mass of the particle is determined from the poles
in the propagators derived from the full Lagrangian \eqleff.
We also plot the one and two loop terms in the coupling
constant expansion of \rhosusy.
The cutoff dependence in the $\rho$ parameter can be seen
from \rhosusy\ to be of order $v^2/\Lambda^2$.\nl
\centerline{\epsfysize=7.0cm\epsfbox{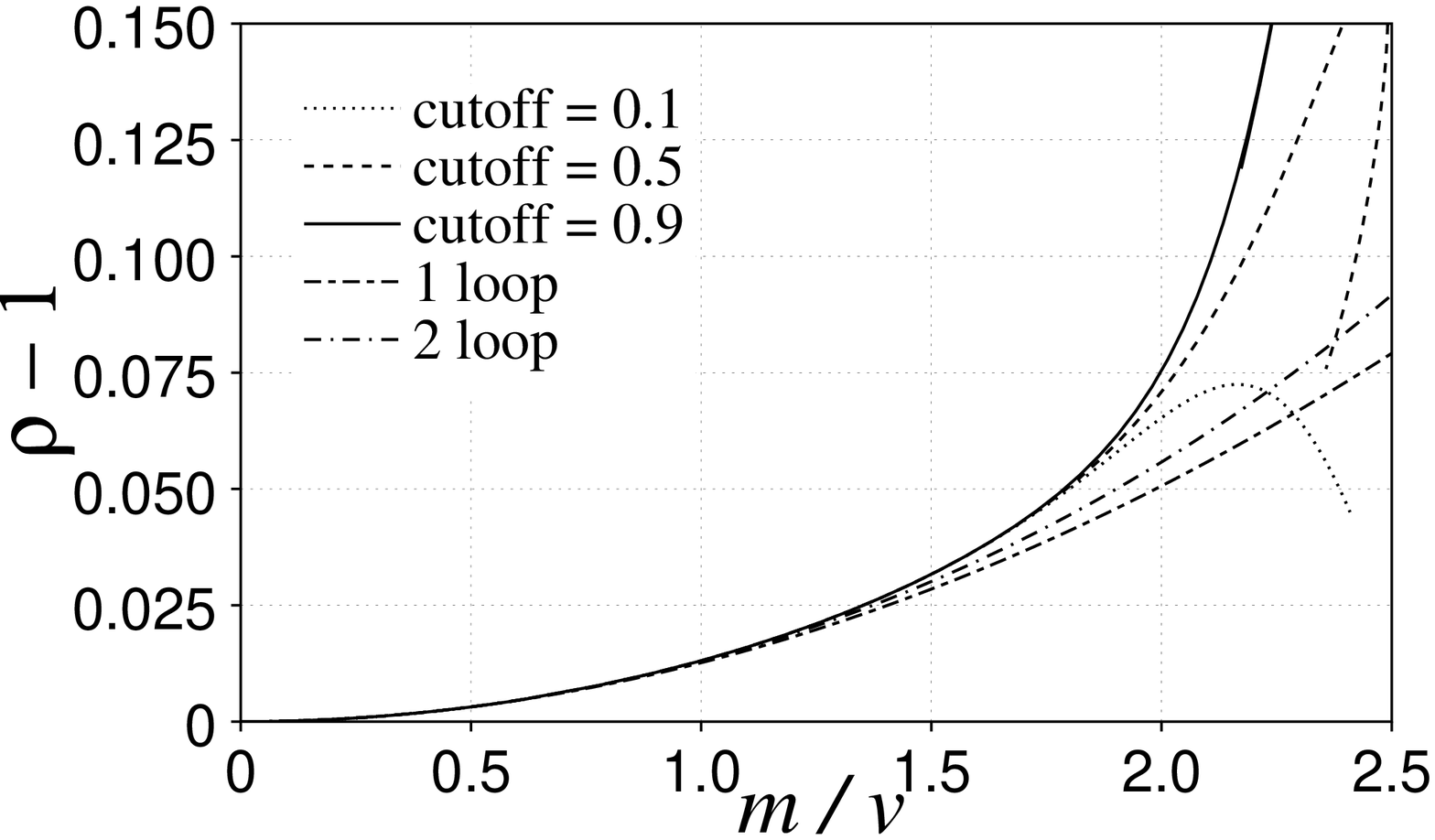}}
\smallskip\noindent
\figsusy{\ \it $\drho$ vs. mass of
$t,\ttl,\ttr$ in units of $v$.}\medskip\noindent
Here and below, we will plot only the region where
the cutoff is larger than the mass.
Also, we will set $\nf=2$ in the plots to put them in
a more familiar context.
For other values of $\nf$, just rescale what we call $v^2$.
The $\rho$ parameter increases with the fermion mass.
Beyond a certain mass ($\sim2v$), the relative size
of the cutoff effects become substantial.
Non--perturbatively, there is a restriction on
the mass of the fermion so that it is smaller than $2.49v$.

If we expand the non--perturbative
expression for the $\rho$--parameter in \rhosusy\ in powers of
the coupling constant, each term in the expansion is finite
without any need for a cutoff.
Lowest order terms in this expansion up to three loop order are
\eqn\srhoexp{
   \drho=2x+4\nf x^2+8\nf^2\left({\pi^2\over3}+1\right)x^3
    +\c O(x^4)
    \qquad\hbox{where \ \ }
 x\equiv{\gf\mtp\over8\sqrt2\pi^2}
}
Here we defined the perturbative mass
$\mtp\equiv y^2\!(\mtp)v^2/2$.

An interesting question we may ask, given
the above results, is how good or how bad perturbation
theory performs in our model.
As $\drho$ increases, the perturbative result becomes
less reliable. However, the cutoff effects that inevitably
arise in the non--perturbative result increases.
As we increase the mass, at some point the non--universal
effects become as large as the deviation of the one loop
result from the non--perturbative result.
Beyond this point, the one loop result is as reliable
(or is unreliable) as the non--perturbative result.
Up to this point, the one loop result deviates from
the non--perturbative result by at most 50\%.
It is also interesting to analyze the significance of the higher
loop contributions:
As the mass increases, the higher loop effects become more
important and they improve the agreement of the
perturbative results with the non--perturbative one.
At some point, however, the non--universal effects become
as large as the improvement the higher loop contribution
makes.
We find that the two loop term contribution is at most 7\%
of $\drho$ before the cutoff effects become as large.
For the three loop contribution, this number is 2\%.
These numbers are products of a rather crude approximation
and cannot be taken too literally;
however, we believe that they illustrate the role
of the perturbative approximation to non--decoupling effects
in a theory with a triviality scale.
The appearance of inherent ambiguities in the non--perturbative
result, of course,
is a limitation of the (supersymmetric) standard model.
Once we have a theory that is well defined in the ultraviolet
that reduces to the standard model in the infrared,
the non--universal effects will disappear.

The $\rho$ parameter may also be computed from the radiative
corrections to the two point functions of
the Nambu--Goldstone bosons as in \rhong.
The result is exactly the same as the one obtained
via the gauge bosons in \rhosusy.
The one particle irreducible contributions to the
two point functions of Nambu--Goldstone bosons $\chiz,\chip$
are listed in \fig\figng{}\nl
\centerline{\epsfxsize=\hsize\epsfbox{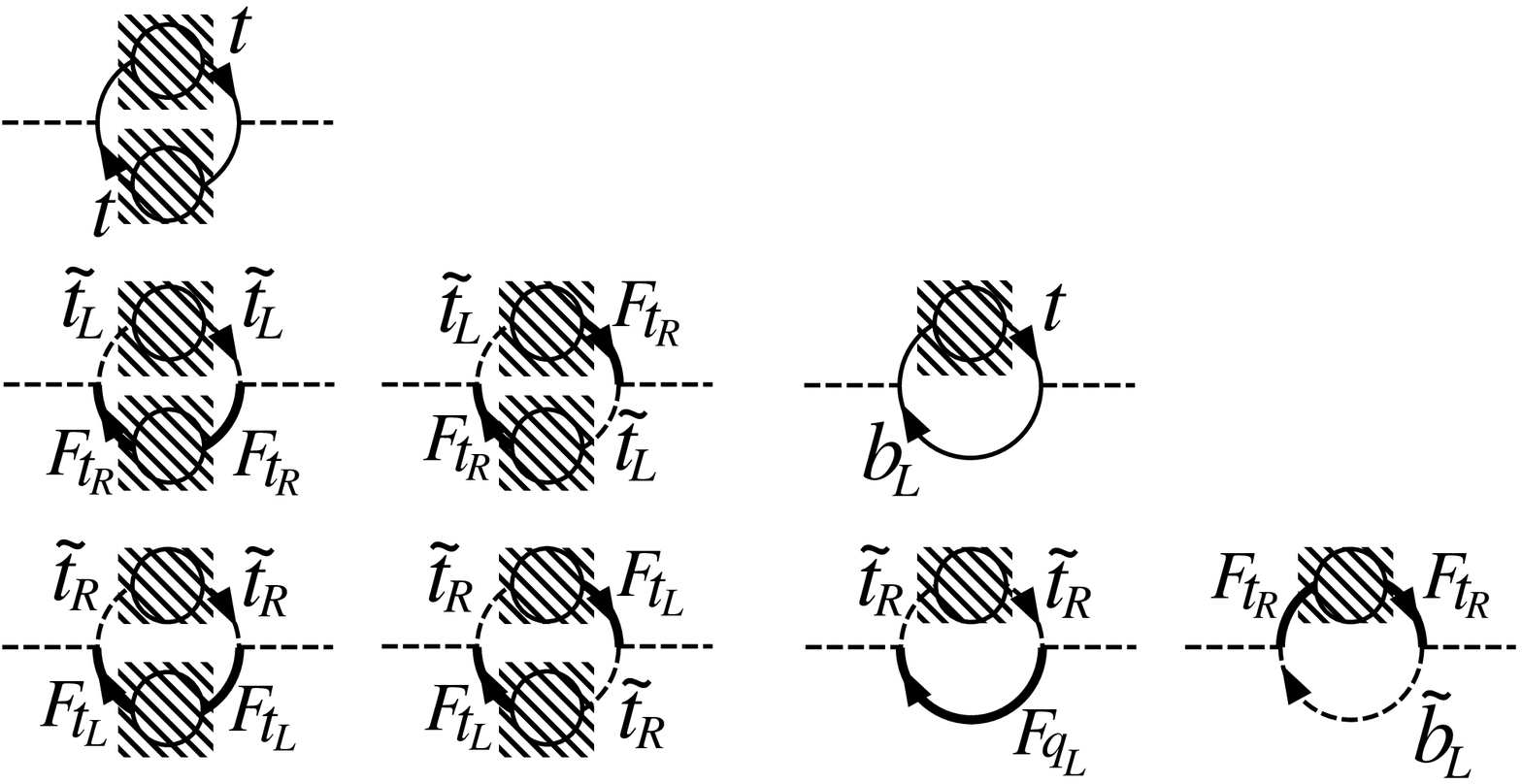}}
\smallskip\noindent\figng{ \it The leading order quantum corrections
to the two point functions of the
Nambu--Goldstone bosons, $\chiz, \chip$.
The left five graphs correct $\chiz$, the right three correct
$\chip$.
}\medskip\noindent
We have used the notations
$\ql^1\equiv\tl,
\tql^1\equiv\ttl,
\Fl^1\equiv\Ftl
$.
We obtain more familiar looking graphs
if we integrate out
the auxiliary fields. However, it seems difficult to directly sum
the contributions corresponding to the graphs in \figng\
once we integrate out the auxiliary fields.

Finally, we note that
the non--perturbative
expression for the $\rho$ parameter \rhosusy\ is identical to
that of the standard model \AP, with the replacement
$\yren\mapsto2\yren$.
Naively, this looks like twice the standard model case,
however we caution that this is not the case, since
the full propagators themselves that enter
the computation contain contributions from the
full supermultiplet.
\newsec{$\rho$--parameter including the effects of the soft supersymmetry
breaking terms}
In this section, we compute the $\rho$--parameter in the
supersymmetric quark--Higgs sector including the effects of the
soft supersymmetry breaking terms.
The $\rho$ parameter may be obtained from the
graphs in \figtwo\ in a calculation similar to that
of the last section, albeit more complicated:
\eqn\rhosb{\drho=
{\hm^4\over v^2}\intk\left[{1\over k^2(\atwo(k^2)k^2+\hm^2)^2}+
{k^2\over(k^2+\mq^2)^2
      \left(\athree(k^2)(k^2+\mq^2)+\hm^2\right)^2}\right]
}
As in the previous section, the renormalized parameters
are used in this expression and $A_i$'s are
the same formulas as $A_{i,bare}$'s in \eqas\
with the replacement,
$y^2\mapsto\yren,\ \sb\mapsto\szero$.
The renormalization scale $\szero$ is arbitrary
and does not affect the results below.

As in the case with no soft breaking terms, this expression
for the $\rho$--parameter requires a cutoff $\Lambda^2$
to give a finite answer. We have plotted the $\rho$
parameter against the physical mass of the fermion, $t$,
for the cutoff values $\Lambda/\sqrt{\striv}=0.1,0.5$ and $0.9$ to
see the cutoff dependence of the expression \rhosb\
in \fig\figsbone{}.
\nl
\centerline{\epsfysize=7.0cm\epsfbox{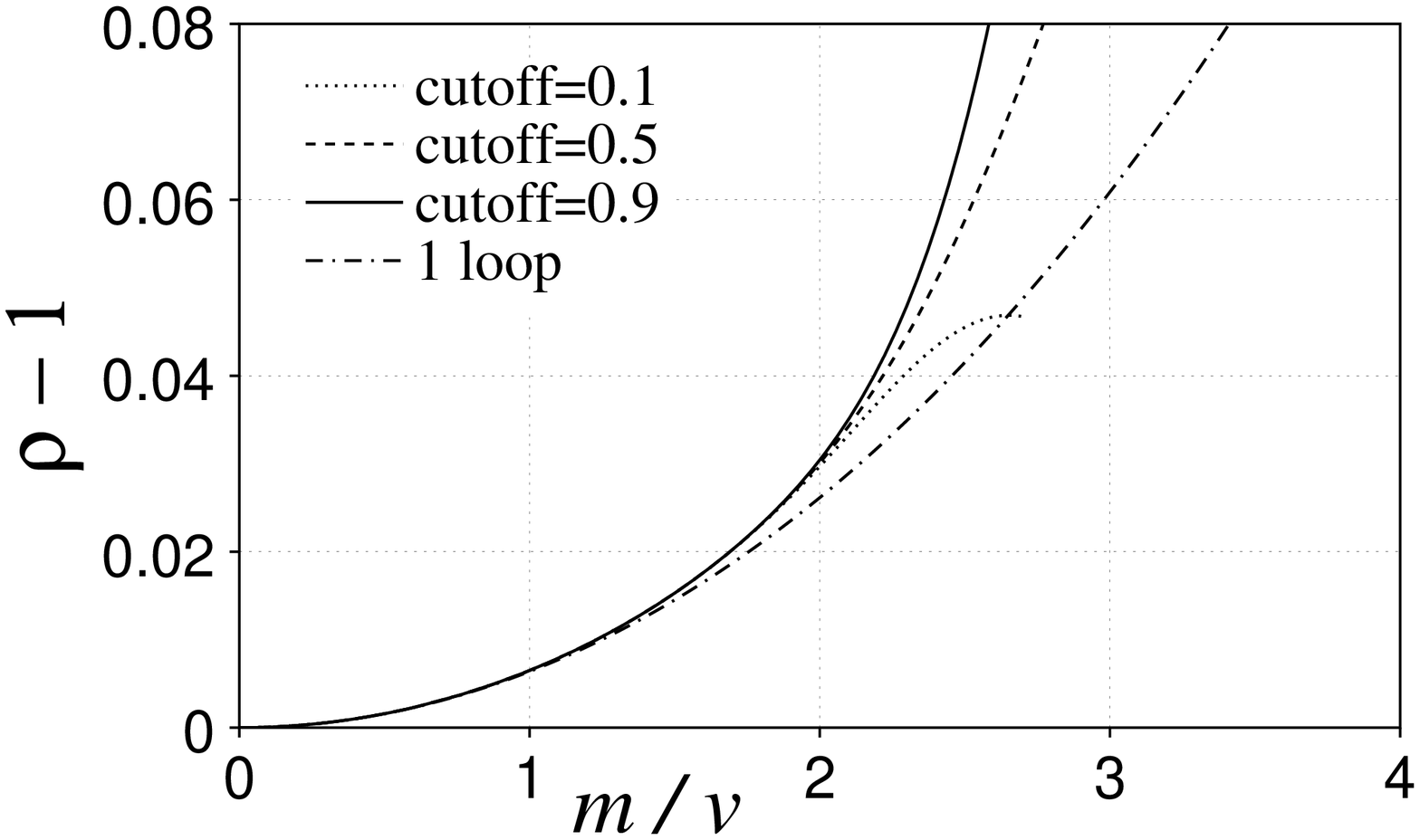}}
\smallskip\noindent
\figsbone{\it\ The parameter $\drho$ against the
mass of $t$  in units of $v$ for the value $\mq^2=(4\pi v)^2/\nf$.}
\medskip\noindent
The soft--breaking scale has been set to $\mq^2=(4\pi v)^2/\nf$
so that its contribution is of the same order as the contribution
from the symmetry
breaking scale.
If we expand the expression in powers of coupling
for $\drho$ in \rhosb,
there is no need to impose a cutoff.
The one loop term in this expansion is
\eqn\rhosboneloop{\drho=2x\left[1+y+y(1+y)\ln{y\over1+y}\right]+\c O(x^2)}
Here $x$ was defined in \srhoexp\ and $y\equiv\mq^2/\hm^2$.
The one loop term, of course, agrees with the previous
literature \SRHO\ and is shown in the plot \figsbone.

The qualitative behavior of the $\rho$ parameter is
the same as the previous section;
as the fermion mass increases, the $\rho$ parameter
also increases.
Beyond a certain mass ($\sim2v$) the non--universal effects become
appreciable.
The maximum fermion mass in this case is $3.58v$.
The non--universal effects in the $\rho$ parameter
may be obtained from \rhosb\ to be of $\c O(v^2/\Lambda^2)$.
It is important to note that the leading order cutoff dependence on $\mq^2$
is further suppressed and is of
$\c O(v^2\mq^2\ln(\mq^2)/\Lambda^4)$.
Therefore the non--universal effects are appreciable
if and only if the fermion mass is large (or the cutoff
is not so high).
In essence, increasing the coupling constant gives rise
to non--decoupling effects and increasing the soft breaking
scale gives rise to decoupling effects.

As in the case without the supersymmetry breaking terms,
we may compute the $\rho$ parameter to leading order
using the propagators for the Nambu--Goldstone bosons
by incorporating the corrections from the graphs in \figng:
\eqn\eqrho{\eqalign{
\drho&={2\hm^4\over v^2}\intk\Biggl[
  {\left(\atwo(k^2)k^2\right)'\over\left(\atwo(k^2)k^2+\hm^2\right)^3}\cr
   &
  +{1\over(k^2+\mq^2)\left(\athree(k^2)(k^2+\mq^2)+\hm^2\right)^2}
   \left(1-{k^2(\athree(k^2)(k^2+\mq^2))'\over
    \athree(k^2)(k^2+\mq^2)+\hm^2}\right)\Biggr]\cr
}}
The primes denote the derivative with respect to $k^2$.
This expression is the same as the expression \rhosb\ obtained
using the current--current correlation function
to all orders in perturbation theory.
The two expressions \rhosb, \eqrho\ differ by cutoff effects
of order $v^2\mq^2/\Lambda^4$,  which are non--perturbative.
This is only to be expected; once the result is non--universal,
physical results are well defined only up to cutoff effects.
Consequently, in general two methods of computation for
the same quantity will yield results which are exactly
the same to all orders in perturbation theory,
but not necessarily the same in the non--universal effects.
(Hence, of course, the name ``non--universal''.)
\newsec{Discussion of the results}
Our objective was to obtain a non--perturbative expression
for a non--decoupling effect, the $\rho$--parameter,
in the supersymmetric standard model and to analyze it.
Considering that the non--decoupling effects become
appreciable only when the coupling becomes large,
this is both a natural and an important problem.
A non--perturbative expression was obtained
including the effects of the soft breaking terms using the
large--$\nf$ expansion.
When we try to evaluate this expression, we find that the expression is
not finite unless we take into account
the fact that the theory, non--perturbatively,
contains an intrinsic cutoff, namely the triviality scale.
Taking this into account and evaluating the expression, we find
that the result is cutoff--dependent when the fermion
mass is large or, equivalently, when the triviality cutoff is small.
That such a seemingly simple closed form expression
is non--perturbative and contains so much physical information,
we find rather remarkable.

Had the cutoff effects appeared for observables measured
at energies close to the cutoff, it would not have been surprising;
what is surprising is that the cutoff effects
appear in a low energy physically measurable
parameter.
The naive expectation for the cutoff effects would
have been of $\c O(p^2/\Lambda^2)$ where $p^2$ is the
momentum scale of interest --- in this case zero.
This naive expectation fails and in fact the
cutoff dependence is of $\c O(v^2/\Lambda^2)$.
This is special to non--decoupling effects;
non--decoupling effects, though
they are measured at low energies, effectively probe
the physics at the electroweak symmetry breaking energy scale,
$v^2$.
Therefore when the fermion mass becomes heavy and the coupling large,
the cutoff becomes of $\c O(v^2)$ so that the cutoff
dependence also becomes appreciable.
In the real world, the cutoff scale corresponds to
the energy scale of new physics that appears beyond
the standard model.
As the mass of the top becomes higher, the energy scale
of new physics becomes lower and the $\rho$--parameter,
along with other non--decoupling parameters, becomes
more sensitive to the new physics that appears.

By contrast, decoupling effects do not see the physics
of the symmetry breaking scale.
The cutoff dependence of the $\rho$ parameter
is of order $v^2/\Lambda^2$
and no contributions of $\c O(\mq^2/\Lambda^2)$ exist.
The cutoff dependence may arise  only when the
coupling is large and the cutoff scale is not so high
compared to the symmetry breaking scale; it cannot
arise when the soft breaking scale is of the
order of the cutoff scale and when $v^2$ is far below it.
Since the soft breaking terms considered here
are not interaction terms,  we expect
that they do  not to affect low energy physics when we increase
it, even to the order of the cutoff.
The results confirm this picture.
However, since the cutoff effects are non--perturbative
in essence, we know of no theorems that apply to this case.

We also analyzed the reliability of the perturbative
approximation to $\drho$ in comparison with the non--perturbative
result.
For instance, we found that the two loop result
improves the one loop result by 10\% at best, since
when the two loop contribution is any larger,
the non--universal effect is at least as large.

A small $\drho$, as seen in nature, may be achieved
by a fermion mass in the perturbative regime.
There is perhaps another intriguing possibility that
the scale of new physics is of order $v^2$ and
the physics beyond the standard model manifests in
the particular form of non--universal effect
so as to make the $\drho$ small.
This would require
very specific properties of higher energy physics and
it is not clear that such a model may be constructed.

Clearly, there is further work to be done:
It would be interesting to study more general models,
such as ones with more complicated Higgs sectors.
Also, it would be interesting to explicitly construct
a model that corresponds to the physics beyond the
triviality cutoff and see how the non--decoupling
effects depend on the particular physics chosen
at the higher energy scales.
The issue of the non--perturbative behavior of non--decoupling
effects is of interest on its own.
To apply this to the standard model ($\nf=2$), $1/\nf$
is {\it a priori }not a small parameter so that it is important to study
the problem using other non--perturbative approaches as well.
An interesting task would be to find a more systematic
understanding of non--decoupling effects in general.

The physics of non--decoupling effects
is qualitatively similar for the supersymmetric and
the non--supersymmetric case.
This is perhaps surprising considering how the quantum
properties of supersymmetric theories markedly differ
from that of non--supersymmetric theories.
However, when we understand the underlying
physics, it is consistent with the behavior of
decoupling effects, non--decoupling
effects and their cutoff dependence.
\medskip\noindent{\bf Acknowledgments: }
We would like to thank Santiago Peris for constructive
comments on the paper and for the pleasant
previous collaboration which lead to this work.
This work was supported in part by the National Science Foundation grant
NSF--89--15286.
\medskip
\listrefs
\end